# Quantum anomalous Hall effect in two-dimensional magnetic insulator heterojunctions

Jinbo Pan[a#], Jiabin Yu[b#], Yan-Fang Zhang[a,c,d], Shixuan Du[c], Anderson Janotti[d], Chao-Xing Liu[b]*, Qimin Yan[a]*

[a] Department of Physics, Temple University, Philadelphia, PA 19122, USA

[b] Department of Physics, the Pennsylvania State University, University Park, PA 16802, USA

[c] Institute of Physics, Chinese Academy of Science, Beijing, 100190, China

[d] Department of Materials Science and Engineering, University of Delaware, DE 19716, USA

Abstract:

**Recent years have witnessed tremendous success in the discovery of topological states of matter. Particularly, sophisticated theoretical methods[1, 2] in time-reversal-invariant topological phases have been developed, leading to the comprehensive search of crystal database and the prediction of thousands of new topological materials[3-5]. In contrast, the discovery of magnetic topological phases that break time reversal is still limited to several exemplary materials because the coexistence of magnetism and topological electronic band structure is rare in a single compound. To overcome this challenge, we propose an alternative approach to realize the quantum anomalous Hall (QAH) effect, a typical example of magnetic topological phase, via engineering two-dimensional (2D) magnetic van der Waals heterojunctions. Instead of a single magnetic topological material, we search for the combinations of two 2D (typically trivial) magnetic insulator compounds with specific band alignment so that they can together form a type-III heterojunction with topologically non-trivial band structure. By combining the data-driven materials search, first principles calculations, and the symmetry-based analytical models, we identify 8 type-III heterojunctions consisting of 2D ferromagnetic insulator materials from a family of 2D**

monolayer MXY compounds (M = metal atoms, X = S, Se, Te, Y = F, Cl, Br, I) as a set of candidates for the QAH effect. In particular, we directly calculate the topological invariant (Chern number) and chiral edge states in the MnNF/MnNCl heterojunction with ferromagnetic stacking. This work illustrates how data-driven material science can be combined with symmetry-based physical principles to guide the search for novel heterojunction-based quantum materials hosting the QAH effect and other exotic quantum states in general.

*Correspondence and requests for materials should be addressed to Q. Y. (qiminyan@temple.edu) and C.X. L. (cxl56@psu.edu)

#J. P. and J. Y. contribute equally to this work.

## Introduction

The quantum anomalous Hall (QAH) effect[6-8], a zero magnetic field manifestation of the integer quantum Hall effect, originates from the exchange interaction between electron spin and magnetism and exhibits quantized Hall resistance and zero longitudinal resistance. Similar to the quantum Hall effect, the QAH effect harbors dissipationless chiral edge states, thus providing an energy-efficient platform for state-of-the-art applications in spintronics[9, 10] and quantum computing[11, 12]. Guided by theoretical predictions[13], the QAH effect was experimentally demonstrated in the magnetically (Cr or V) doped $(Bi,Sb)_2Te_3$[14-17]. However, magnetic doping reduces the spin-orbit coupling strength and thus may drive the system into a trivial phase. It also degrades the sample quality by bringing a large amount of disorder scatterings into the system. Consequently, the critical temperature of the QAH state through the magnetic doping approach is usually below 2 K, an order-of-magnitude lower than the Curie temperature of these compounds. Therefore, it is highly desirable to search for new platforms consisting of stoichiometric materials with intrinsic magnetism to realize high temperature QAH state.

Recent rapid development of two-dimensional (2D) layered magnetic insulators has provided an exciting new opportunity for the realization of QAH effect in un-doped material systems. The QAH effect has been proposed to appear in the heterostructures of topological insulator (TI) and magnetic insulator through the magnetic proximity effect across the interface[18-22]. Another recent advance is the successful synthesis of a class of intrinsic magnetic TI materials, including $MnBi_2Te_4$, $MnBi_2Se_4$ and $MnSb_2Te_4$[23, 24], in which A-type antiferromagnetism and topologically non-trivial band structure coexist, leading to the successful observation of the QAH state[25-27]. The QAH effect has also been observed in twisted bilayer graphene[28, 29].

Despite several recent works[30, 31], the search for more QAH materials is still challenging because it requires the coexistence of ferromagnetism and topological band structure, which is rare in a single material. In this letter, we propose an alternative strategy to realize the QAH effect in 2D magnetic van der Waals (VdW) heterojunctions by combining data-driven discovery of novel 2D magnetic compounds with a theoretical analysis of topological properties. All the previous works[18-21, 30, 31] on the QAH heterojunctions require at least one intrinsic topological material. In contrast, every QAH VdW heterojunction predicted by our approach can be constructed with two trivial 2D ferromagnetic (FM) insulators. The key idea is to make sure the two FM 2D materials form the so-called broken gap junction or type-III junction, in which the conduction-band minimum (CBM) of one 2D material is lower in energy than the valence-band maximum (VBM) of the other compound at certain high-symmetry momentum, as shown in Fig. 1(a). In other words, an inverted (topological) band structure is formed by the heterojunction, rather than one individual material. Since this strategy only requires the (valence or conduction) band offset to be larger than the band gap of one of the two 2D materials, it greatly relaxes the stringent conditions for the QAH effect and allows us to develop a systematic approach to identify the candidate compounds for the QAH heterojunctions in 2D magnetic VdW material database.

This approach has been successfully utilized to search for the quantum spin Hall insulators, such as InAs/GaSb quantum wells[32] and other semiconductor heterojunctions[33], but it has not been applied to the search of QAH systems, mainly due to the lack of 2D magnetic materials database. Therefore, a data-driven computational and theoretical search of 2D magnetic materials is called for. Fig. 1(b) summarizes the workflow for our data-driven computational approach, which is carefully designed to incorporate high-throughput calculations at multiple computational levels and symmetry-based physical principles that are critical to quickly identify the most promising candidates hosting the QAH effect. In this initial work, we focus on a family of 2D monolayer MXY compounds (M = metal atoms, X = S, Se, Te, Y = F, Cl, Br, I) and predict 46 potential 2D magnetic semiconductors in this family. A database including the information of CBM, VBM, band offset, the orbital nature and the symmetry properties of the conduction and valence bands, and the magnetic anisotropy energies and exchange coupling for each candidate compound is created. Based on this 2D magnetic material database, we identify 8 VdW heterojunctions made from different combination of these compounds to possess inverted band structure.

We next perform a generic symmetry analysis of topological properties for the VdW heterojunctions. This analysis is based on the relationship between topological invariant, the Chern number (CN) that characterizes the QAH effect, and the two-fold rotation ($C_2$) eigenvalues[1, 2, 34-36]. We demonstrate that the CN for the heterojunction must be changed by an odd number when the magnetizations in two magnetic VdW compounds are switched from FM to anti-ferromagnetic stacking order. This suggests that the QAH phase must exist for at least one magnetic stacking order.

Finally, we choose MnNF/MnNCl heterojunctions as an example and construct a tight-binding model that captures the main features of the band structure from the realistic material simulations. Based on this tight-binding model, we explicitly compute the CN and the chiral edge states of this heterojunction. Our results show that the QAH phase occurs for the FM stacking order of the magnetization between two materials of the heterojunction and is consistent with our symmetry analysis by computing the $C_2$ eigenvalues from first principles calculations.

**Prediction of 2D material insulators in Monolayer MXY compounds**

In the past decade, a tremendous amount of computed and experimental material data has been shared in the materials research community through multiple material databases for both bulk[37-39] and 2D inorganic compounds[40-44], This offered a vast inorganic materials space that is explorable by data-driven approaches and enabled the data-driven discovery of novel quantum materials hosting exotic topological phases[4, 5, 30, 31, 45]. Here our theoretical predictions of 2D heterojunctions that host the QAH effect is motivated by the data-driven discovery of a family of 2D monolayer MXY compounds (M = metal atoms, X = S, Se, Te, Y = F, Cl, Br, I) that are predicted to host both FM and nonmetallic ground states. We initiated our material discovery efforts from a material set of around 560 monolayer MXY compounds that are created through elemental substitution based on the existing MXY materials in public 2D materials databases[40-44]. The data-driven materials discovery process is illustrated in Fig. 1(b) and more computational details can be found in the Fig. S1 of the Supplemental Information (SI). With FM state as the initial magnetic configuration, through an energy minimization and electron density optimization process, we identified 46 potential magnetic semiconductors (see Table S1 in the SI). 2D FM insulators MnNX and CrCX (X = Cl, Br, I; C = S, Se, Te) have been predicted in previous work[46-49]. Our comprehensive study puts this 2D compound family as a great platform for the discovery and design of quantum materials based on 2D magnetic heterojunctions.

From 10 different FM MXY compounds, we propose 8 VdW heterojunctions, in which the CBM of one 2D layer (called conduction layer) is below the VBM of the other 2D layer (called valence layer). Thus, all these heterojunctions exhibit the desirable band inversion, as well as small-enough (< 6.5%) lattice mismatch, as shown in Table 1. The Curie temperatures ($T_c$) of the these 10 monolayer FM MXY compounds are evaluated by performing Monte Carlo (MC) simulations based on a 2D Heisenberg Hamiltonian model[50, 51]. The interaction types, the values of exchange coupling $J$ and magnetic anisotropy energy $A$ are summarized in Fig. 1(a) and Table 2. Computational details related to magnetic anisotropy energies and exchange coupling interactions are included in the SI. Characterized by high Curie temperature as well as small band gaps, this set of magnetic semiconductor compounds provides a great platform for the design of van der Waals heterojunctions to realize the QAH effect. The CBM of all the conduction layers and the VBM of all the valence layers are both located at $\Gamma$ point, facilitating the analysis of type-III junctions.

**Symmetry analysis of topological properties for the VdW heterojunctions**

We next analyze the topological properties of the proposed 8 VdW heterojunctions based on the symmetry aspect. The magnetic moments are in the $z$ direction (normal to the 2D planes) for all these magnetic MXY compounds, which reduce their plane groups to $p2$, generated by $C_2$ and lattice translations. Putting two 2D materials together to form a heterojunction can preserve the $p2$ symmetry as long as the magnetic moments of two layers are in the same or opposite directions (named as FM or AFM stacking, respectively). Using $C_2$ eigenvalues[1, 2, 34-36], we will argue that the two stacking orders result in different CNs once the heterojunctions are in the inverted regime, meaning that at least one way gives QAH heterojunctions.

Owing to $C_2^2 = -1$, the states at four $C_2$-invariant momenta have definite $C_2$ eigenvalues $\pm i$, as exemplified in Fig. 3. The quantity of interest is the total number $n_-$ of occupied states with $C_2$ eigenvalue $-i$ at all $C_2$-invariant momenta, since

$$CN \bmod 2 = n_- \bmod 2 \qquad (1)$$

holds for a 2D insulator[34, 35]. To get a clear physical picture, we first neglect the spin-orbit coupling (SOC) and the interlayer coupling between two layers in the VdW heterojunctions but include the charge transfer that equalizes the chemical potentials of the two layers. Such approximation allows us to treat spin and layer indexes as good quantum numbers (see Figs. 3(a-b)), and then we can split $n_-$ into $n_- = n_{-,v} + n_{-,c}$ with $n_{-,c}$ and $n_{-,v}$ from the conduction and valence layers, respectively. The only difference between the FM and AFM stacking is the relative direction of magnetic moments in the two constituting compounds. Without loss of generality, we reverse the direction of the magnetic moment of the conduction layer when switching the stacking order. Since the magnetization in the valence layer stays fixed, we have $n_{-,v}^{FM} = n_{-,v}^{AFM}$ with the superscripts indicating the stacking order.

To address $n_{-,c}$, we further split $n_{-,c}$ into $n_{-,c} = n_{-,c,low} + n_{-,c,high}$, where $n_{-,c,low}$ comes from the band that is partially filled (involved in the band inversion) and $n_{-,c,high}$ is given by the occupied bands fully below $E_F$ (see Figs. 3(a-b)). In the conduction layer, the sign of $C_2$ eigenvalue of any state is flipped when switching from the FM stacking to the AFM stacking, as the orbital nature is unchanged while the spin direction is reversed. It results in $n_{-,c,high}^{AFM} = 4N_{c,high} - n_{-,c,high}^{FM}$, where the pre-factor 4 comes from the four $C_2$-invariant momenta and $N_{c,high}$ is the number of occupied bands fully below $E_F$ in the conduction layer. Since only one band from the conduction layer is partially filled and it only contributes to $n_{-,c}$ at $\Gamma$ for the heterojunctions of interest, $n_{-,c,low}$ is either 0 or 1, leading to $n_{-,c,low}^{AFM} = 1 - n_{-,c,low}^{FM}$. In sum, we have $n_-^{AFM} = n_-^{FM} + 1 + 4N_{c,high} - 2n_{-,c}^{FM}$, resulting in

$$n_-^{AFM} \bmod 2 = n_-^{FM} + 1 \bmod 2. \qquad (2)$$

Now we discuss the SOC and the interlayer coupling. The energy scales of these two effects are typically much smaller than the energy scale of the band inversion (*i.e.*, the energy gap at $\Gamma$ in this case), which is determined by the hopping and the magnetic exchange coupling. Indeed, as shown by the results from first-principle calculations (see SI), the SOC and interlayer coupling do not cancel the existing band inversion or introduce new band inversion in the heterostructures of interest, leaving Eq. (2) unchanged. However, the two effects are still important for our analysis since they open the gap at generic momenta (not $C_2$-invariant momenta) and enable us to combine Eq. (2) with Eq. (1). (See schematic examples in Figs. 3(c-d).) As a result, we conclude that the parity of the CN is changed when switching the stacking order of the 2D heterojunction (from FM to AFM or the inverse), and thus at least one of the stacking orders can give non-zero CN.

The remaining question is which stacking order guarantee a nonzero CN for the heterojunctions of interest. Let us assume that the CNs of the two individual materials have the same parity, e.g., both materials have zero CNs, so that the total $n_-$ of the two materials must be even before forming the heterojunction. Then, the key lies on the $C_2$ eigenvalues of the two states involved in the band inversion at $\Gamma$, *i.e*, CBM of the conduction layer and VBM of the valence layer in this case. Specifically, if the two states have opposite $C_2$ eigenvalues for a certain stacking order, $n_-$ of the heterojunction must differ by 1 from the total $n_-$ of the two materials before forming the heterojunction (See Fig. 3(a)). Thus, such stacking order must yield a QAH heterojunction according to Eq. (1). With this principle in mind, we list the $C_2$ eigenvalues of the VBM and

CBM for all FM materials with all magnetic moments fixed in the positive z direction (see Table. S2 in SI). From this list, we can directly read out the $C_2$ eigenvalues of the two states involved in the band inversion at the Γ point for any heterojunction with FM stacking, and if they are opposite (same), the heterojunction has non-zero CN for the FM (AFM) stacking.

**The QAH effect in MnNF/MnNCl heterojunction**

Due to the fact that all the 10 2D FM materials that we identified in this class have magnetization moments along the z direction and have *p2* symmetry, the discussion above based on symmetry eigenvalues is valid for all the 8 van der Waals heterojunctions identified. To conclusively demonstrate the emergence of QAH effect from these 2D magnetic VdW heterojunctions, we focus on a prototype system consisting of two magnetic compounds, MnNF and MnNCl with small band gaps. Both compounds are predicted to host FM ground state. The Curie temperatures of these two compounds are evaluated by both HSE06 and PBE+U functionals. We observe that hybrid function predicts lower transition temperatures than PBE+U for MnNF and MnNCl, owing to the fact that the inclusion of exact exchange in HSE06 functional generates more localized states on Mn atoms and weaker exchange interactions among them. Even in this case, the predicted Curie temperatures for MnNF and MnNCl are 671 K and 592 K, respectively (Fig. 2(a)), well above room temperature. Both partial density of states and orbital-projected band structures of MnNF and MnNCl (Fig. 2(b)) indicate that the conduction band edges of these compounds are mainly d states while the valence band edge is derived from N $p_y$ states.

Based on the calculated band alignment between the two compounds (as shown in Fig. 2(c)), we expect that a type-III heterojunction can be formed withband inversion. As shown in Fig. 2(d), we create MnNF/MnNCl VdW heterojunctions with both FM and AFM stacking. Atomic structures are relaxed using HSE06. Considering that VdW interaction is not included in the hybrid functional HSE06, we artificially apply a set of small strain along *z*-direction for the MnNF/MnNCl heterojunction and find that the band structure is robust against strain. In the case of FM stacking, without SOC, the crossing between two bands in the same spin channel happens around Γ point, leading to the band inversion (Fig. 2(e)). The inclusion of SOC opens band gaps at the original crossing points of the two bands, which are located at certain generic momenta. In the AFM stacking case, at the Fermi level the band overlap happens between the two bands from two different spin channels.

In MnNF and MnNCl, within the *p2mm* plane group, Mn cations are confined in a quasi-octahedral bonding environment associated with a $t_{2g}$ - $e_g$ d state splitting. This crystal field splitting (1~2 eV) provides an essential electronic structure scaffold for small-gap FM semiconductors. Orbital analysis confirms that the CBM of MnNF or MnNCl is composed of Mn $e_g$ states, while the VBM is derived from the interaction between occupied N *p* states and Mn $t_{2g}$ states. This unique electronic structure feature endows the VBM and CBM of MXY compounds different $C_2$ rotational eigenvalues (see Table S3 in the SI) that is essential for the emergence of QAH effect in their heterojunctions. From our first-principles calculations, the CBM and VBM at the Γ point (zone center) of MnNF and MnNCl have opposite $C_2$ eigenvalues in the FM configuration, indicating that the heterojunction with FM stacking must have non-zero CN according to the above symmetry analysis. To better understand the result of first-principle calculation, we construct effective models around Γ point near the Fermi energy. Near $E_F$, the conduction and valence bands of both materials are composed of $p_y$ and $d_{x^2-y^2}$ orbitals, respectively (as shown in Figs. 2(b) and 2(c)). For the FM stacking, the basis can be chosen as $(|d_{x^2-y^2}, \uparrow\rangle, |p_y, \uparrow\rangle)$, resulting in the effective model:

$$\begin{pmatrix} E_v + B_{vx}k_x^2 + B_{vy}k_y^2 & -iA_y k_y + \lambda_1 k_x \\ iA_y k_y + \lambda_1 k_x & E_c + B_{cx}k_x^2 + B_{cy}k_y^2 \end{pmatrix}. \tag{3}$$

Here $\lambda_1$ is given by SOC, all other parameters are dominated by the hopping. More details are provided in the SI. According to the band structure from first principles calculations, we may choose $E_v - E_c > 0$, $B_{vx} - B_{cx} < 0$, and $B_{vy} - B_{cy} < 0$. Then, the direct calculation of the CN yields $CN_{FM}^{eff} = -sgn(A_y \lambda_1)$, which is non-zero. On the other hand, the effective model for AFM stacking has the form:

$$\begin{pmatrix} E_v + B_{vx}k_x^2 + B_{vy}k_y^2 & -i\lambda_0 \\ i\lambda_0 & E_c + B_{cx}k_x^2 + B_{cy}k_y^2 \end{pmatrix}. \tag{4}$$

with bases $(|d_{x^2-y^2}, \uparrow\rangle, |p_y, \downarrow\rangle)$ and $\lambda_0$ induced by SOC. The *k*-independent off-diagonal term leads to CN equals zero. Since the occupied bands fully below $E_F$ have the same CNs for different stacking orders, the total CNs should differ by one, confirming the above conclusion.

We then construct a tight-binding model for further clarification of the topological properties of the heterostructure. The basis of the model is given by the $d_{x^2-y^2}$, $d_{xy}$ and $d_{yz}$ orbitals of the Mn atoms and the $p_y$ orbital of the N atoms. The FM of each layer is included by adding the localized magnetic moments on the Mn atoms and the on-site SOC is also included. With reasonable parameter values, the bulk band structures of the FM and AFM stacked heterostructures are shown in Fig. 4(a) and (b), respectively, which qualitatively match those obtained by the first-principle calculations (see the SI for more details). Direct calculations for the FM and AFM stacked heterostructure yield CN = 1 and CN = 0, respectively, coinciding with the chiral edge mode for FM stacking in Fig. 4(c) and the fully gapped edge for AFM stacking in Fig. 4(d). The tight-binding model calculation verifies different topological properties of the FM and AFM stacked heterostructures.

**Conclusion and Discussion**

In summary, we proposed a new approach to realize the QAH effect in 2D magnetic VdW heterojunctions and conclusively demonstrate the existence of the QAH effect in MnNF/MnNCl heterojunction with FM stacking by combining the first principles calculations and the symmetry analysis. We also proposed other VdW heterojunctions that can host the QAH effect, as listed in Table 1, and the FM stacking always guarantees non-zero CN for them if each individual material is topologically trivial. It is important to note that Mn, Cr, Mo, and W atoms in these compounds that form VdW heterojunctions are all in a $d^3$ electronic configuration with $t_{2g}$ states occupied and $e_g$ states empty. Given the high Curie temperatures (above room temperature) of many compounds in the MXY family (see Table 2), these VdW heterojunctions provide an ideal platform to explore the possibility of room-temperature QAH state. The electronic structures and topological properties of VdW heterojunctions can be effectively controlled by strain and electric field (see Fig. S5). Moreover, the design principles proposed in this work is not limited to this family of materials and can be extended/applied to a large variety of 2D magnetic systems in multiple plane groups protected by various symmetry operations, which will greatly expand the family of heterojunction-based QAH systems in the future. Therefore, our work will stimulate experimental synthesis and measurement efforts in this class of heterojunction systems and accelerate the discovery and design of new QAH materials, as well as novel systems with other magnetic topological phases.

**Methods**

*First-principles calculations based on density functional theory (DFT).* In our calculations, geometric relaxations and total energy calculations were performed using density functional theory within projector-augmented wave (PAW) potentials[52, 53] as implemented in the VASP code[54]. A vacuum slab of 20 Å and a plane-wave basis set with an energy cutoff of 520 eV were used. A 10×8×1 Γ-point centered k-point was applied to sample the Brillouin zone. For the initial steps of discovery process, GGA+U was employed to optimize the geometric structures[55], where the U values of transition metal atoms are outlined in Wang et al.'s work[56]. The structures were fully relaxed until energy and force were converged to $10^{-6}$ eV and 0.01 eV/Å, respectively. The screened hybrid functional of Heyd, Scuseria, and Ernzerhof (HSE06)[57, 58] was used to compute the band structures of MXY monolayers. For the 10 promising FM MXY scaffolding compounds, HSE06 calculations were performed to optimize the geometric structures, and compute the magnetic interactions. To verify the dynamical stability of monolayer MnNF and MnNCl, phonon dispersion analysis was performed by using the density functional perturbation theory method as implemented in the Phonopy code[59], interfaced with VASP. In phonon calculations, GGA+U along with the projector-augmented wave (PAW) potentials were employed, with a convergence criterion of $10^{-8}$ eV for energy. For the ab initio molecular dynamics (MD) simulations[60, 61], a canonical (NVT) ensemble was used. Time interval between each step is 1 fs.

*Note added*: During the final stage of this work, a related work appeared on arXiv which performed a systematic search for magnetic topological materials based on magnetic topological quantum chemistry approach[62].

**Acknowledgments**

We thank Cui-zu Chang, Xiaodong Xu for the helpful discussion. J. Pan and Q. Yan acknowledge support from the U.S. Department of Energy, Office of Science, Basic Energy Sciences, under Award #DE-SC0019275 for the design of data-driven discovery pipeline and the first-principles computational work. J.B. Yu and C.X. Liu acknowledge the support of DOE grant (DE-SC0019064) for the analytical model and symmetry analysis, and also the Office of Naval Research (Grant No. N00014-18-1-2793), as well as Kaufman New Initiative research grant of the Pittsburgh Foundation. A. Janotti acknowledges support from U.S. DOE SE-SC0014388. S.X. Du thanks the International Partnership Program of Chinese Academy of Sciences, Grant number No. 112111KYSB20160061. It benefitted from the supercomputing resources of the National Energy Research Scientific Computing Center (NERSC), a U.S. Department of Energy Office of Science User Facility operated under Contract No. DE-AC02-05CH11231, and Temple University's HPC resources supported in part by the National Science Foundation through major research instrumentation grant number 1625061 and by the US Army Research Laboratory under contract number W911NF-16-2-0189.

**References:**

1. Po, H. C.; Vishwanath, A.; Watanabe, H. *Nat Commun* **2017,** 8, (1), 50.
2. Bradlyn, B.; Elcoro, L.; Cano, J.; Vergniory, M. G.; Wang, Z.; Felser, C.; Aroyo, M. I.; Bernevig, B. A. *Nature* **2017,** 547, (7663), 298-305.
3. Vergniory, M. G.; Elcoro, L.; Felser, C.; Regnault, N.; Bernevig, B. A.; Wang, Z. *Nature* **2019,** 566, (7745), 480-485.
4. Zhang, T.; Jiang, Y.; Song, Z.; Huang, H.; He, Y.; Fang, Z.; Weng, H.; Fang, C. *Nature* **2019,** 566, (7745), 475-479.
5. Tang, F.; Po, H. C.; Vishwanath, A.; Wan, X. *Nature* **2019,** 566, (7745), 486-489.
6. Liu, C.-X.; Zhang, S.-C.; Qi, X.-L. *Annual Review of Condensed Matter Physics* **2016,** 7, (1), 301-321.
7. Wang, J.; Lian, B.; Zhang, S.-C. *Physica Scripta* **2015,** T164, 014003.
8. Chang, C.-Z.; Li, M. *Journal of Physics: Condensed Matter* **2016,** 28, (12), 123002.
9. Wu, J.; Liu, J.; Liu, X.-J. *Physical review letters* **2014,** 113, (13), 136403.
10. Zhang, R.-X.; Hsu, H.-C.; Liu, C.-X. *Phys Rev B* **2016,** 93, (23), 235315.
11. Chen, C.-Z.; Xie, Y.-M.; Liu, J.; Lee, P. A.; Law, K. T. *Physical Review B* **2018,** 97, (10), 104504.
12. Zeng, Y.; Lei, C.; Chaudhary, G.; MacDonald, A. H. *Phys Rev B* **2018,** 97, (8), 081102.
13. Yu, R.; Zhang, W.; Zhang, H.-J.; Zhang, S.-C.; Dai, X.; Fang, Z. *Science* **2010,** 329, (5987), 61-64.
14. Chang, C.-Z.; Zhao, W.; Kim, D. Y.; Zhang, H.; Assaf, B. A.; Heiman, D.; Zhang, S.-C.; Liu, C.; Chan, M. H. W.; Moodera, J. S. *Nat Mater* **2015,** 14, (5), 473-477.
15. Chang, C.-Z.; Zhang, J.; Feng, X.; Shen, J.; Zhang, Z.; Guo, M.; Li, K.; Ou, Y.; Wei, P.; Wang, L.-L.; Ji, Z.-Q.; Feng, Y.; Ji, S.; Chen, X.; Jia, J.; Dai, X.; Fang, Z.; Zhang, S.-C.; He, K.; Wang, Y.; Lu, L.; Ma, X.-C.; Xue, Q.-K. *Science* **2013,** 340, (6129), 167-170.
16. Checkelsky, J.; Yoshimi, R.; Tsukazaki, A.; Takahashi, K.; Kozuka, Y.; Falson, J.; Kawasaki, M.; Tokura, Y. *Nature Physics* **2014,** 10, (10), 731.
17. Bestwick, A. J.; Fox, E. J.; Kou, X.; Pan, L.; Wang, K. L.; Goldhaber-Gordon, D. *Phys Rev Lett* **2015,** 114, (18), 187201.
18. Li, M.; Chang, C.-Z.; Kirby, B. J.; Jamer, M. E.; Cui, W.; Wu, L.; Wei, P.; Zhu, Y.; Heiman, D.; Li, J.; Moodera, J. S. *Phys Rev Lett* **2015,** 115, (8), 087201.
19. Tang, C.; Chang, C.-Z.; Zhao, G.; Liu, Y.; Jiang, Z.; Liu, C.-X.; McCartney, M. R.; Smith, D. J.; Chen, T.; Moodera, J. S.; Shi, J. *Science Advances* **2017,** 3, (6), e1700307.
20. Katmis, F.; Lauter, V.; Nogueira, F. S.; Assaf, B. A.; Jamer, M. E.; Wei, P.; Satpati, B.; Freeland, J. W.; Eremin, I.; Heiman, D.; Jarillo-Herrero, P.; Moodera, J. S. *Nature* **2016,** 533, (7604), 513-516.
21. Fu, H.; Liu, C.-X.; Yan, B. **2019,** *arXiv preprint arXiv*:1908.04322.
22. Zou, R.; Zhan, F.; Zheng, B.; Wu, X.; Fan, J.; Wang, R. **2020,** *arXiv preprint arXiv*:2002.12624.
23. Li, J.; Li, Y.; Du, S.; Wang, Z.; Gu, B.-L.; Zhang, S.-C.; He, K.; Duan, W.; Xu, Y. *Science Advances* **2019,** 5, (6), eaaw5685.
24. Otrokov, M. M.; Klimovskikh, I. I.; Bentmann, H.; Estyunin, D.; Zeugner, A.; Aliev, Z. S.; Gaß, S.; Wolter, A. U. B.; Koroleva, A. V.; Shikin, A. M.; Blanco-Rey, M.; Hoffmann, M.; Rusinov, I. P.; Vyazovskaya, A. Y.; Eremeev, S. V.; Koroteev, Y. M.; Kuznetsov, V. M.; Freyse, F.; Sánchez-Barriga, J.; Amiraslanov, I. R.; Babanly, M. B.; Mamedov, N. T.; Abdullayev, N. A.; Zverev, V. N.; Alfonsov, A.; Kataev, V.; Büchner, B.; Schwier, E. F.; Kumar, S.; Kimura, A.;

Petaccia, L.; Di Santo, G.; Vidal, R. C.; Schatz, S.; Kißner, K.; Ünzelmann, M.; Min, C. H.; Moser, S.; Peixoto, T. R. F.; Reinert, F.; Ernst, A.; Echenique, P. M.; Isaeva, A.; Chulkov, E. V. *Nature* **2019,** 576, (7787), 416-422.
25.	Liu, C.; Wang, Y.; Li, H.; Wu, Y.; Li, Y.; Li, J.; He, K.; Xu, Y.; Zhang, J.; Wang, Y. **2019**, *arXiv preprint arXiv*:1905.00715.
26.	Deng, Y.; Yu, Y.; Shi, M. Z.; Guo, Z.; Xu, Z.; Wang, J.; Chen, X. H.; Zhang, Y. *Science* **2020,** 367, (6480), 895-900.
27.	Ge, J.; Liu, Y.; Li, J.; Li, H.; Luo, T.; Wu, Y.; Xu, Y.; Wang, J. **2019**, *arXiv preprint arXiv*:1907.09947.
28.	Serlin, M.; Tschirhart, C. L.; Polshyn, H.; Zhang, Y.; Zhu, J.; Watanabe, K.; Taniguchi, T.; Balents, L.; Young, A. F. *Science* **2020,** 367, (6480), 900-903.
29.	Sharpe, A. L.; Fox, E. J.; Barnard, A. W.; Finney, J.; Watanabe, K.; Taniguchi, T.; Kastner, M. A.; Goldhaber-Gordon, D. *Science* **2019,** 365, (6453), 605.
30.	Choudhary, K.; Garrity, K. F.; Jiang, J.; Pachter, R.; Tavazza, F. **2020**, *arXiv preprint arXiv*:2001.11389.
31.	Olsen, T.; Andersen, E.; Okugawa, T.; Torelli, D.; Deilmann, T.; Thygesen, K. S. *Physical Review Materials* **2019,** 3, (2), 024005.
32.	Liu, C.; Hughes, T. L.; Qi, X.-L.; Wang, K.; Zhang, S.-C. *Phys Rev Lett* **2008,** 100, (23), 236601.
33.	Zhang, H.; Xu, Y.; Wang, J.; Chang, K.; Zhang, S.-C. *Phys Rev Lett* **2014,** 112, (21), 216803.
34.	Turner, A. M.; Zhang, Y.; Mong, R. S. K.; Vishwanath, A. *Phys Rev B* **2012,** 85, (16), 165120.
35.	Hughes, T. L.; Prodan, E.; Bernevig, B. A. *Phys Rev B* **2011,** 83, (24), 245132.
36.	Kruthoff, J.; de Boer, J.; van Wezel, J.; Kane, C. L.; Slager, R.-J. *Physical Review X* **2017,** 7, (4), 041069.
37.	Bergerhoff, G.; Hundt, R.; Sievers, R.; Brown, I. *Journal of chemical information and computer sciences* **1983,** 23, (2), 66-69.
38.	Jain, A.; Ong, S. P.; Hautier, G.; Chen, W.; Richards, W. D.; Dacek, S.; Cholia, S.; Gunter, D.; Skinner, D.; Ceder, G. *Apl Materials* **2013,** 1, (1), 011002.
39.	Gražulis, S.; Daškevič, A.; Merkys, A.; Chateigner, D.; Lutterotti, L.; Quiros, M.; Serebryanaya, N. R.; Moeck, P.; Downs, R. T.; Le Bail, A. *Nucleic acids research* **2011,** 40, (D1), D420-D427.
40.	Ashton, M.; Paul, J.; Sinnott, S. B.; Hennig, R. G. *Physical review letters* **2017,** 118, (10), 106101.
41.	Choudhary, K.; Kalish, I.; Beams, R.; Tavazza, F. *Scientific reports* **2017,** 7, (1), 5179.
42.	Haastrup, S.; Strange, M.; Pandey, M.; Deilmann, T.; Schmidt, P. S.; Hinsche, N. F.; Gjerding, M. N.; Torelli, D.; Larsen, P. M.; Riis-Jensen, A. C. *2D Materials* **2018,** 5, (4), 042002.
43.	Mounet, N.; Gibertini, M.; Schwaller, P.; Campi, D.; Merkys, A.; Marrazzo, A.; Sohier, T.; Castelli, I. E.; Cepellotti, A.; Pizzi, G. *Nature nanotechnology* **2018,** 13, (3), 246.
44.	Zhou, J.; Shen, L.; Costa, M. D.; Persson, K. A.; Ong, S. P.; Huck, P.; Lu, Y.; Ma, X.; Chen, Y.; Tang, H. *Scientific data* **2019,** 6, (1), 86.
45.	Vergniory, M.; Elcoro, L.; Felser, C.; Regnault, N.; Bernevig, B. A.; Wang, Z. *Nature* **2019,** 566, (7745), 480.
46.	Guo, Y.; Zhang, Y.; Yuan, S.; Wang, B.; Wang, J. *Nanoscale* **2018,** 10, (37), 18036-18042.

47.     Qi, J.; Wang, H.; Qian, X. **2018**, *arXiv preprint arXiv:*1811.02674.
48.     Wang, C.; Zhou, X.; Zhou, L.; Tong, N.-H.; Lu, Z.-Y.; Ji, W. *Science Bulletin* **2019,** 64, (5), 293-300.
49.     Jiang, Z.; Wang, P.; Xing, J.; Jiang, X.; Zhao, J. *ACS applied materials & interfaces* **2018,** 10, (45), 39032-39039.
50.     Xiang, H.; Lee, C.; Koo, H.-J.; Gong, X.; Whangbo, M.-H. *Dalton Transactions* **2013,** 42, (4), 823-853.
51.     Xiang, H.; Kan, E.; Wei, S.-H.; Whangbo, M.-H.; Gong, X. *Physical Review B* **2011,** 84, (22), 224429.
52.     Blöchl, P. E. *Physical review B* **1994,** 50, (24), 17953.
53.     Kresse, G.; Joubert, D. *Physical Review B* **1999,** 59, (3), 1758.
54.     Kresse, G.; Furthmüller, J. *Computational materials science* **1996,** 6, (1), 15-50.
55.     Dudarev, S. L.; Botton, G. A.; Savrasov, S. Y.; Humphreys, C. J.; Sutton, A. P. *Phys Rev B* **1998,** 57, (3), 1505-1509.
56.     Wang, L.; Maxisch, T.; Ceder, G. *Phys Rev B* **2006,** 73, (19), 195107.
57.     Heyd, J.; Scuseria, G. E.; Ernzerhof, M. *J Chem Phys* **2006,** 124, (21), 219906.
58.     Heyd, J.; Scuseria, G. E.; Ernzerhof, M. *J Chem Phys* **2003,** 118, (18), 8207-8215.
59.     Togo, A.; Tanaka, I. *Scripta Materialia* **2015,** 108, 1-5.
60.     Kresse, G.; Hafner, J. *Phys Rev B* **1993,** 47, (1), 558-561.
61.     Kresse, G.; Hafner, J. *Phys Rev B* **1994,** 49, (20), 14251-14269.
62.     Xu, Y.; Elcoro, L.; Song, Z.; Wieder, B. J.; Vergniory, M. G.; Regnault, N.; Chen, Y.; Felser, C.; Bernevig, B. A. **2020**; *arXiv preprint arXiv*:2003.00012.

Table 1. Potential van der Waals heterojunctions based on ferromagnetic MXY compounds with electronic band inversions and lattice mismatches less than 6.5%. "Nontrivial stacking" indicates the stacking order that guarantees a nonzero CN under the assumption that each individual material is topologically trivial. Valence-band maximum (VBM) and conduction-band minimum (CBM) are given with respect to vacuum level.

| Material A | CBM (eV) | Band Character (CBM) | Material B | VBM (eV) | Band Character (VBM) | Lattice mismatch along $a$ (%) | Lattice mismatch along $b$ (%) | Nontrivial Stacking |
|---|---|---|---|---|---|---|---|---|
| FeNF | -7.9 | Fe: $d_{x^2-y^2}$ | MnNF | -7.6 | N: $p_y$ | 2.3 | 1.1 | FM |
| VSI | -5.9 | V: $d_{xy}$ | MoSI | -5.6 | Mo: $d_{yz}$ | 1.7 | 3.2 | FM |
| VSI | -5.9 | V: $d_{xy}$ | WSeI | -4.8 | W: $d_{yz}$ | 3.9 | 6.1 | FM |
| VSI | -5.9 | V: $d_{xy}$ | VSeI | -5.6 | Se: $p_x$ / I: $p_x$ | 2.7 | 6.0 | FM |
| MnNF | -7.1 | Mn: $d_{x^2-y^2}$ | MnNCl | -6.7 | N: $p_y$ | 6.5 | 0.6 | FM |
| TiTeI | -4.9 | Ti: $d_{z^2}$ | WTeI | -4.9 | W: $d_{yz}$ | 4.0 | 3.0 | FM |
| VSeI | -5.1 | V: $d_{z^2}$ / $d_{x^2-y^2}$ | WSeI | -4.8 | W: $d_{yz}$ | 6.5 | 0.1 | FM |
| CrSeCl | -5.6 | Cr: $d_{x^2-y^2}$ | VSeI | -5.6 | Se: $p_x$ / I: $p_x$ | 4.1 | 0.5 | FM |

Table 2: Exchange coupling parameter $J$ (in meV), magnetic anisotropy energy $A$ (in meV), and estimated Curie temperature Tc of MXY compounds based on HSE06 calculations.

|  | $J_{12}$ | $J_{13}$ | $J_{34}$ | $A_{[100]}$ | $A_{[010]}$ | $A_{[001]}$ | Tc (K) |
|---|---|---|---|---|---|---|---|
| TiTeI | 9.33 | -13.19 | -1.22 | -5.0 |  | 0 | 46 |
| FeNF | 0.09 | -19.85 | -8.90 | -0.01 |  | 0 | 398 |
| VSI | -93.41 | -73.03 | -43.59 | 12.93 |  | 0 | 1100 |
| VSeI | -7.03 | -34.49 | -24.83 | 0.02 | -1.90 | 0 | 913 |
| MnNF | -16.39 | -8.73 | -2.46 | 0.34 |  | 0 | 671 |
| MnNCl | -16.13 | -5.67 | -2.91 | 0.37 |  | 0 | 592 |
| CrSeCl | -3.79 | -6.13 | 1.69 | 0.62 |  | 0 | 270 |
| MoSI | 1.07 | -3.08 | 1.07 | 0.02 |  | 0 | 76 |
| WSeI | 4.03 | -7.68 | -6.38 | 0.03 |  | 0 | 302 |
| WTeI | 2.12 | -11.86 | -6.26 | 0.00 | -0.25 | 0 | 479 |

Figure 1. Illustration of a data-driven materials discovery and design framework that combines high-throughput computations and symmetry-based analysis for the search of heterojunction-based 2D magnetic topological systems that host the QAH effect. (a) Schematic of hypothesis: QAH effect that can be emerged from type-III 2D VdW heterojunctions. (b) Data-driven discovery pipeline for VdW-heterojunction-based QAH systems.

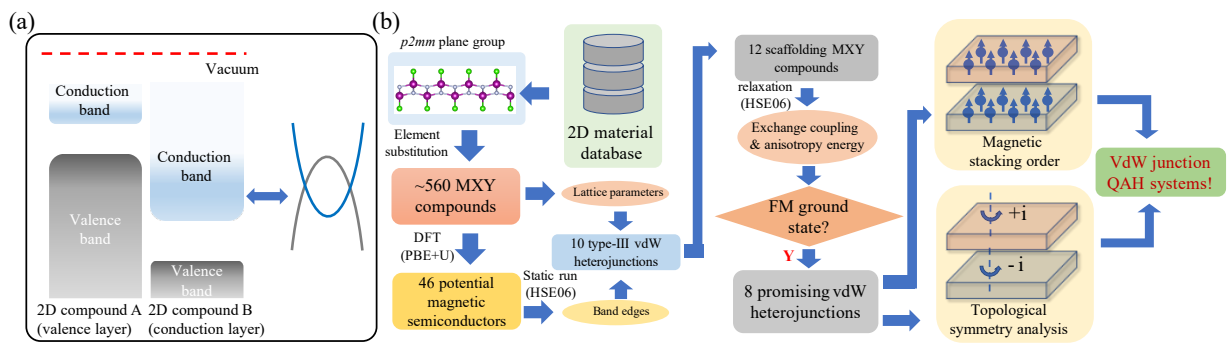

Figure 2. Computational data for a representative material system: MnNF, MnNCl, and their VdW heterojunctions. (a) Estimation of Curie temperatures for MnNF and MnNCl by Monte-Carlo simulations based on a Heisenberg Hamiltonian model that is fitted to both HSE06 and PBE+U results; (b) and (c) Orbital projected band structures of MnNF and MnNCl; (d) The relaxed structure of MnNF/MnNCl heterojunction; (e) Band structures (with SOC) of the FM-stacking and (f) AFM-stacking MnNF/MnNCl heterojunctions.

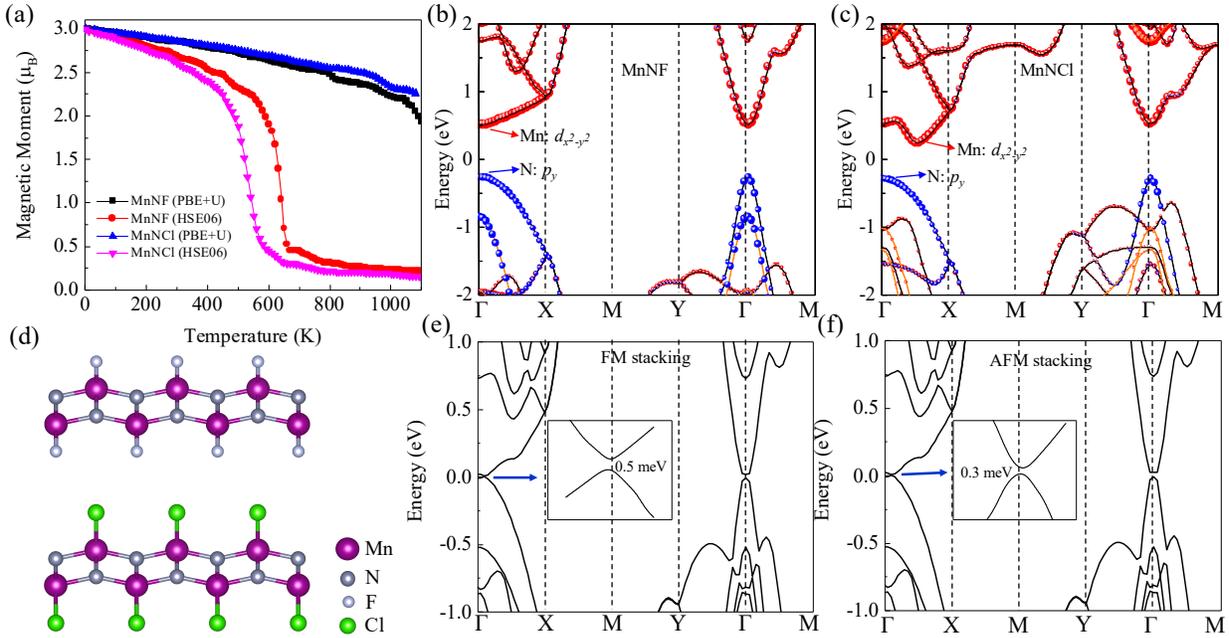

Figure 3. Schematics of the examples with certain $n_-$. "IC" labels the interlayer coupling and the black dashed line is the Fermi energy. $\pm$ correspond to the $C_2$ eigenvalues $\pm i$, with the black labeling the occupied states and the orange labeling the empty states involved in the band inversion at $\Gamma$. In (a-b), there are no SOC and no interlayer coupling, the blue and red lines come from the valence and conduction layers, respectively, and ↑↓ stand for the spin polarization, whereas SOC and the interlayer coupling are included in (c-d). (a) and (c) are for FM stacking with $n_-^{FM} = 5$, where CN must be non-zero in (c) and the band inversion happens between two states with opposite $C_2$ eigenvalues. Flipping the spins in the conduction layer gives AFM stacking with $n_-^{AFM} = 8$ as shown in (b) and (d). In this case, CN might be zero and the band inversion happens between two states with the same $C_2$ eigenvalues.

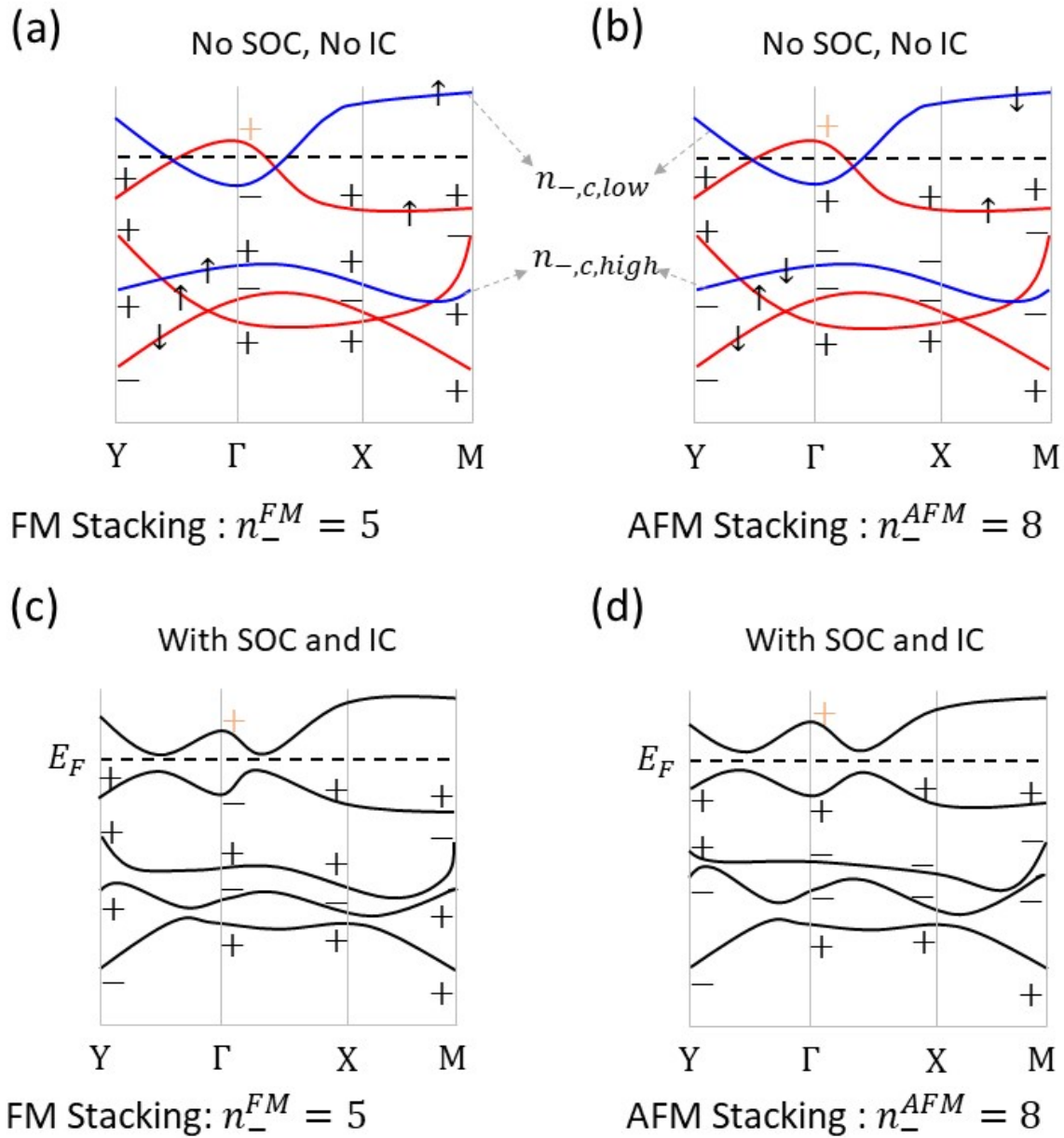

Figure 4. Electronic band structures generated by the tight-binding model. (a-b) show the bulk band structure, and (c-d) illustrate the energy dispersion along the (01) edge. In (c-d), $a$ is the lattice constant along $x$ direction.

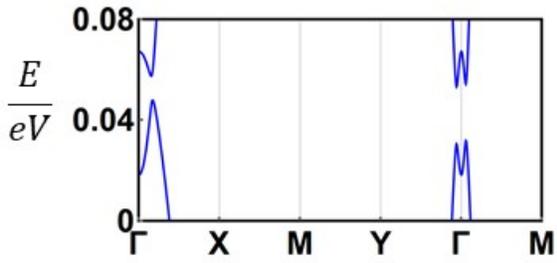
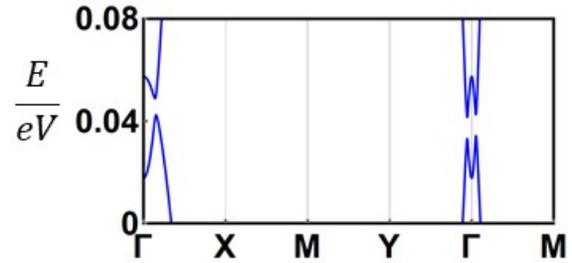
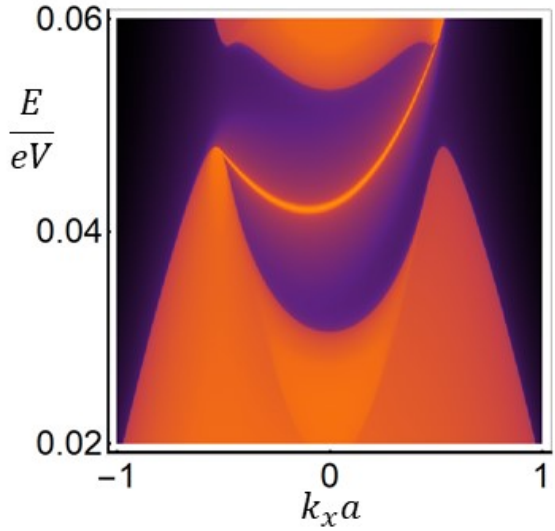
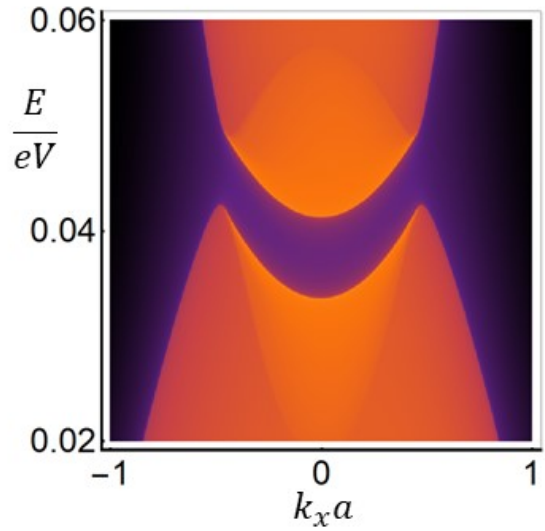